\documentclass[final,3p,times,twocolumn]{elsarticle}

\usepackage{amssymb}
\usepackage{graphicx}
\usepackage{natbib}

\usepackage{lineno}

\begin{document}
\begin{frontmatter}



\title{Front-end electronics and data acquisition system for a multi-wire 3D gas tracker}


\author{K. {\L}ojek\fnref{adr1}}
\author{D. Rozp\c{e}dzik\corref{cor1}\fnref{adr1}}
\ead{dagmara.rozpedzik@uj.edu.pl}
\author{K. Bodek\fnref{adr1}}
\author{M. Perkowski\fnref{adr1,adr2}}
\author{N. Severijns\fnref{adr2}}

\address[adr1]{Institute of Physics, Jagiellonian University, 30348 Cracow, Poland}
\address[adr2]{Instituut voor Kern-en Stralingsfysica, University of Leuven, 3001 Leuven, Belgium}
\cortext[cor1]{Corresponding author}


\begin{abstract}
This paper presents the design and implementation of the front-end electronics and the data acquisition 
(DAQ) system for readout of multi-wire drift chambers (MWDC). Apart of the conventional drift time measurement 
the system delivers the hit position along the wire utilizing the charge division technique. The system consists 
of preamplifiers, and analog and digital boards sending data to a back-end computer via an Ethernet interface. 
The data logging software formats the received data and enables an easy access to the data analysis software. 
The use of specially designed preamplifiers and peak detectors allows the charge-division readout of the low resistance signal wire. 
The implication of the charge-division circuitry onto the drift time measurement was studied and the overall performance of the electronic
 system was evaluated in dedicated off-line tests.
\end{abstract}


\end{frontmatter}


\section{Introduction}
Precision measurements in neutron and nuclear decay offer a sensitive window to search for 
new physics beyond the standard electroweak model and allow also the determination of the fundamental 
weak vector coupling. Recent analyses based on the effective field theory performed in e.g.~\cite{1,2}
show that in processes involving the lightest quarks the neutron and nuclear decay will compete with 
experiments at highest energy accelerators. For instance, data taken at the LHC is currently probing 
these interactions at the $10^{-2}$ level (relative to the standard weak interactions), with the potential to 
reach the $\simeq10^{-3}$ level. In some of the $\beta$ decay correlation measurements there are prospects to reach experimental
 sensitivities between $10^{-3}$ and $10^{-4}$ making these observables interesting probes for searches of new physics 
originating at TeV scale. The most direct access to the exotic tensor interaction in $\beta$ decay is to measure the 
Fierz term (coefficient $b$) or the beta-neutrino correlation coefficient $a$ in a pure Gamow-Teller transition~\cite{3}. 
The $b$ coefficient shows up as a tiny energy dependent $(1/E)$ departure of the $\beta$ spectrum from its V-A (standard model) shape.
 The smallness of the potential $b$ contribution requires that other corrections to the spectrum shape of the same order are 
included in the analysis. Indeed, according to~\cite{4,5} the recoil terms also affect the spectrum shape with their main contribution
 being proportional to $E$. In order to disentangle these effects the detector efficiency for $\beta$ particle as a function of energy must be known with
 the precision better than $10^{-3}$~\cite{6}. The dominating contribution
 in the systematic uncertainty comes from back-scattering and out-scattering of electrons from the detector. Monte Carlo simulation of
 this effect is helpful, however, it introduces its own uncertainty as the input parameters are known with limited accuracy. 
Monte Carlo simulation would reflect the real situation better after it is adjusted to real experimental data of a particular measurement setup. 
 
 The described in this article electronics was designed for a spectrometer capable of direct registration of the back-scattering events, 
thus providing reference data for the Monte Carlo calculation of the detector efficiency. The spectrometer itself is still in an R\&D phase
 undergoing detailed tests and tuning. It will be a subject of a separate paper together with the performance benchmark~\cite{7}. In this paper, its concept
 will be described only in a minimum extent at the beginning of Section 2 to explain the requirements imposed onto the front-end electronics and DAQ.
 The rest of Section 2 is devoted to the electronic system architecture. The test results were obtained with a help of a signal generator and are presented in Section 3.
 Therein the resulting time spectrum and the charge asymmetry distribution representing the typical performance are shown. The asymmetry spectra were obtained using a
 dedicated tester to simulate the hit position on the wire with adjustable resistance division (potentiometer). Conducting the electronic benchmark tests without a detector
 was chosen by purpose. The detector itself is still not fully understood. Therefore it was important to assess the purely electronic contribution to the performance 
parameters of the spectrometer. The paper ends with a short summary and outlook for the future experiment.

\section{System architecture}
\begin{figure}[!t]
\centering
 \includegraphics[scale=0.31]{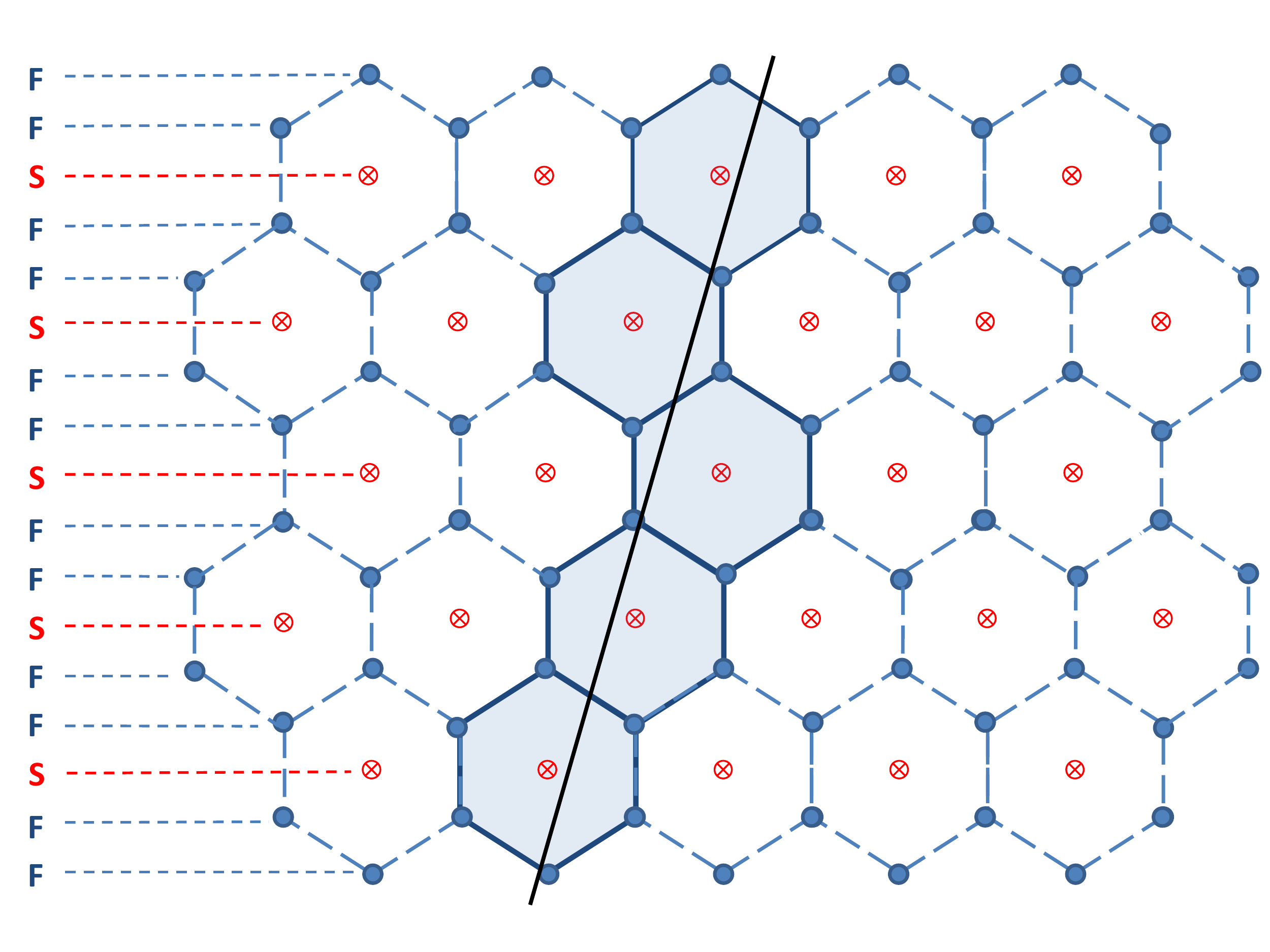}
 \caption{(color on-line) Fragment of the hexagonal wire structure. The blue (solid) circles denote the field wires and the red circles (crossed) correspond to signal wires. 
 The field and signal wire planes are indicated with F and S, respectively. 
The shaded areas ilustrate the responding cells after an electron passed along the black solid line.}
\label{cells}
\end{figure}

In order to facilitate the identification of the electrons impinging onto and scattered from the energy detector (e.g. Si detector, scintillator)
 a low-Z and low-mass tracker must be applied. One of the attractive options is a low pressure multi-wire drift chamber with minimum number of necessary wires
in order to maximize the detector transparency. This condition can be fulfilled by a hexagonal wire geometry and the charge division technique allowing for a 3D track 
reconstruction without major distortion of
 the electron energy measurement. The hexagonal wire geometry is not the only one considered in the project. The rectangular (planar) wire configuration is the
 next suitable alternative. 
The multi-wire drift chamber is based on the small prototype described in Ref.~\cite{8} and 
will be operated with He/Isobutan gas mixture (ranging from 70\%/30\% to 90\%/10\%) at lowered pressure (down to 300 mbar).
 It consists of 10 sense wire planes (8 wires for each plane) separated with 24 field wire planes forming the double F-F-S-F-F-S-F-F-S-F-F-S-F-F-S-F-F structure
 (F- denotes a field wire plane, S- denotes a signal wire plane). The distance between neighboring signal planes is 15 mm, with wires within a plane being separated 
by 17.32 mm. This wire plane structure leads to the hexagonal cell geometry as shown in Fig.~\ref{cells}.
 Each cell consists of a very thin anode wire (NiCr alloy, 25 $\mu$m diameter) with resistance of about 20 ohms/cm surrounded by 6 cathode field wires forming
 a hexagonal cuboid. All wires are soldered to pads of the printed circuit board (PCB) frames. The chamber is equipped with a two-dimensional positioning system
 for a beta source installed in the central region of the detector, between both parts of the MWDC structure. In the initial configuration, the electrons detected in 
two plastic scintillators installed at both sides of the MWDC provide the time reference signal for the drift time measurement. The PMT signals are also used as a trigger 
for the MWDC and electron energy detector readout. Acquiring the drift time and the pulse height asymmetry at both ends of the responding signal wire one can establish the 
electron path across the chamber cell. The system provides charge-division position sensing in the direction parallel to the wires as well as precise drift time measurement. 
The expected position resolution across wires is limited by angular straggling of primary electrons travelling in the gas and accounts to about 450 $\mu$m as shown in Ref.~\cite{8}. 
In this situation, the precision of the drift time measurement of about 200 ps is more than needed as it corresponds to about 50 $\mu$m for the operating conditions in view. 
The MWDC will be operated in homogenous magnetic field oriented
 parallel to wires providing a rough electron transverse momentum filter. The position information along the wires will be used for the identification of the sequence 
of the cells passed by the electrons and for distinguishing between the electrons impinging onto and scattered from the energy detector. This is why the modest position
 resolution of a few mm is sufficient for that direction.

Crucial in the design is the analog part of the system. In principle, commercial digitizers (TDC, ADC) could be applied at the end of the acquisition chain.
However, it has been decided to equip the data channels with custom digital boards with adjusted specifications such that the whole system 
is handy, cost effective and scalable. The applied on board processing (FPGA) allows for acquiring up to 15 000 triggered events per second which gives a 
comfortable factor of 10 reserve as compared to the application in view.      

\begin{figure}[!t]
\centering
 \includegraphics[scale=0.45]{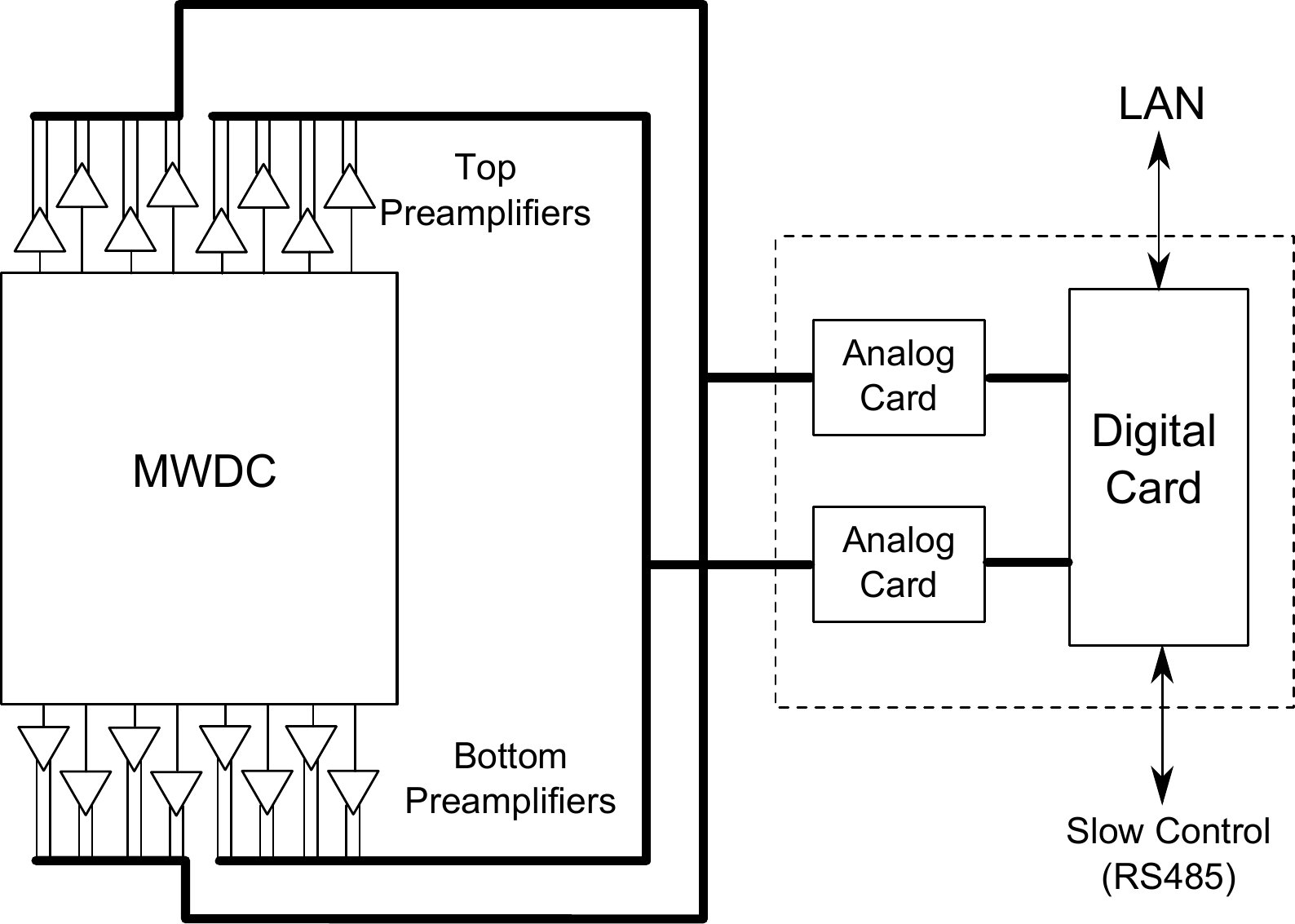}
 \caption{Schematic of the data acquisition architecture.}
\label{daq}
\end{figure}

\begin{figure}[t]
\begin{center}
 \includegraphics[scale=0.2]{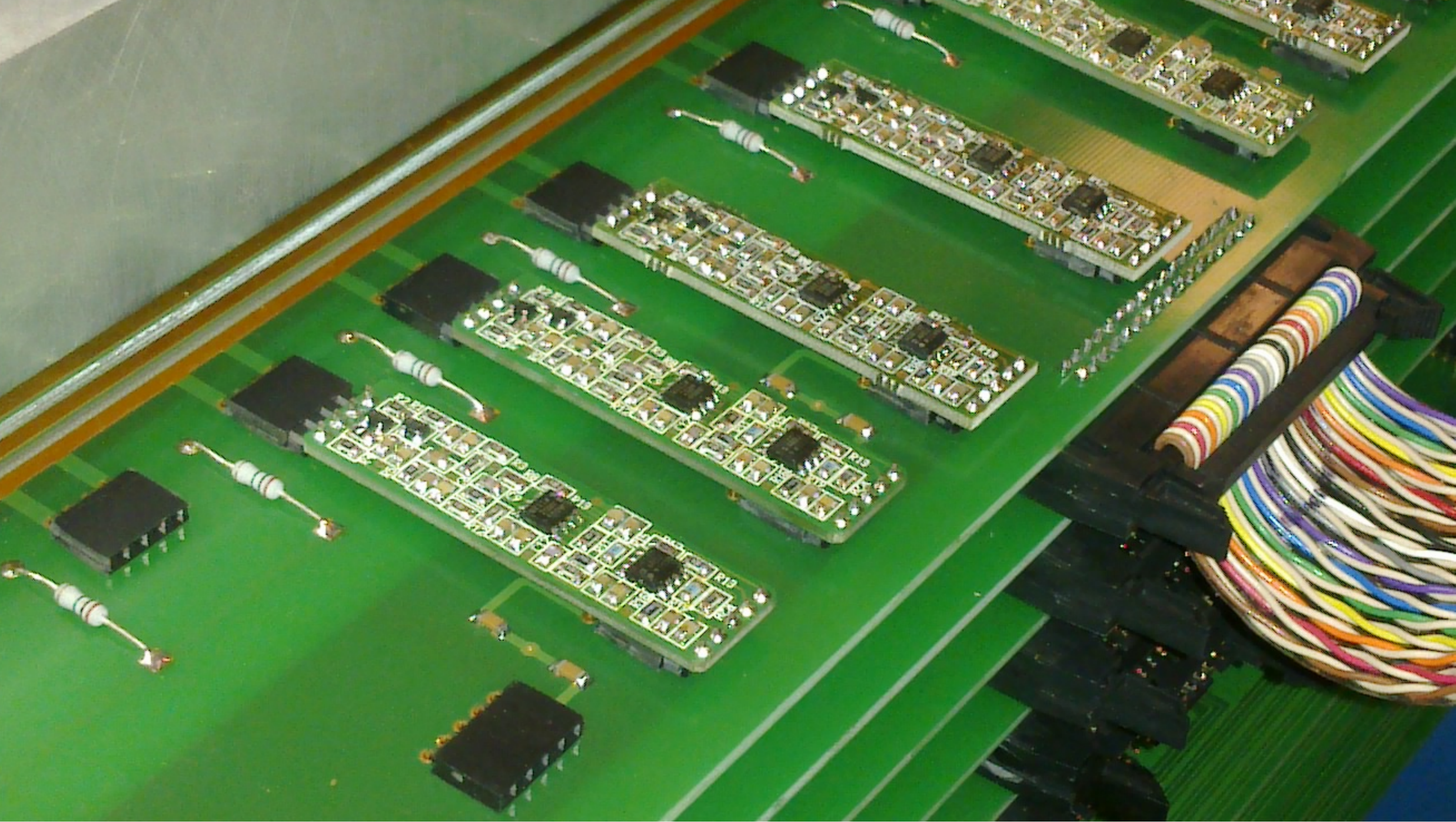}
 \caption{(color on-line) Small piggyback preamplifier PCB cards plugged into the wire frames.}
\label{pcbpre}
\end{center}
\end{figure}

The described modular electronic system consists of three main parts: (i) preamplifiers, (ii) analog cards containing the peak
 detector and constant fraction discriminator (CFD), and (iii) the digital boards containing analog to digital converters, ADC and TDC. 
The data logging software on the PC constitutes the data receiver. The signal from each signal plane is processed by means of one electronics module,
 which consists of two analog cards and one digital board providing 16 ADC and 8 TDC channels. A corresponding block diagram is shown in Fig.~\ref{daq}.
The signals from both ends of a signal wire are fed to inputs of preamplifiers located directly on the detector frames (see Fig.~\ref{pcbpre}) in order to minimize the input noise
 and protect the signal from EM interferences. The signals from the preamplifiers are received by the analog cards which drive the ADC inputs and produce the
 TDC STOP signals. The digital data is transmitted via a LAN port to the back-end computer where the process of receiving, sorting and formatting to a complete physical
 event is accomplished by the data logging software. The card configuration settings and control is done via a RS485 port. 
Detailed description of each part of the system is presented in the following sections.

\subsection{Preamplifiers}
\begin{figure}[!t]
\centering
 \includegraphics[scale=0.29]{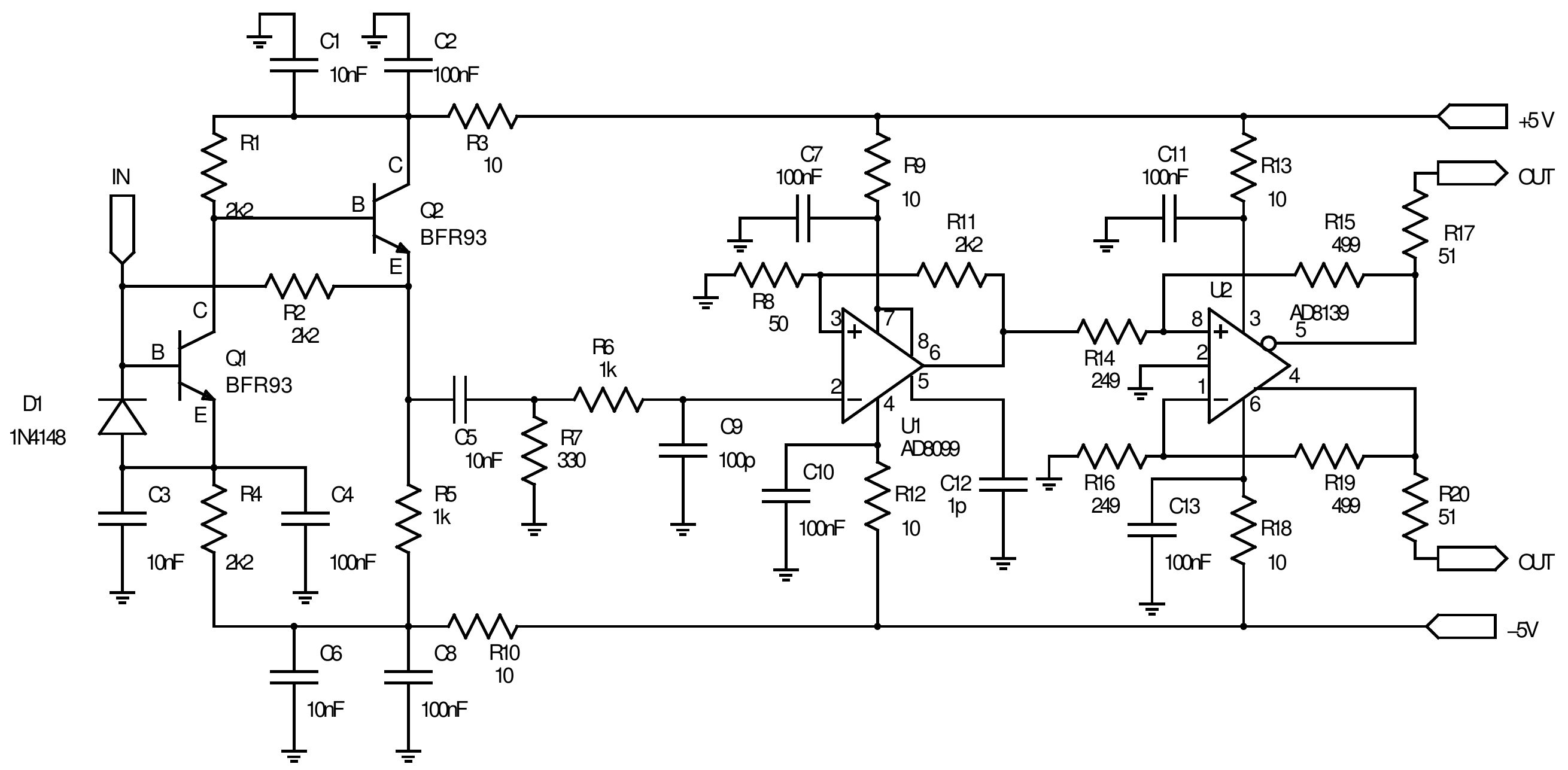}
 \caption{Electronic circuit of the preamplifier. The system uses 160 dedicated preamplifiers to the drift chamber signals with charge division.}
\label{preamp}
\end{figure}

\begin{figure}[!t]
\begin{center}
 \includegraphics[scale=0.6]{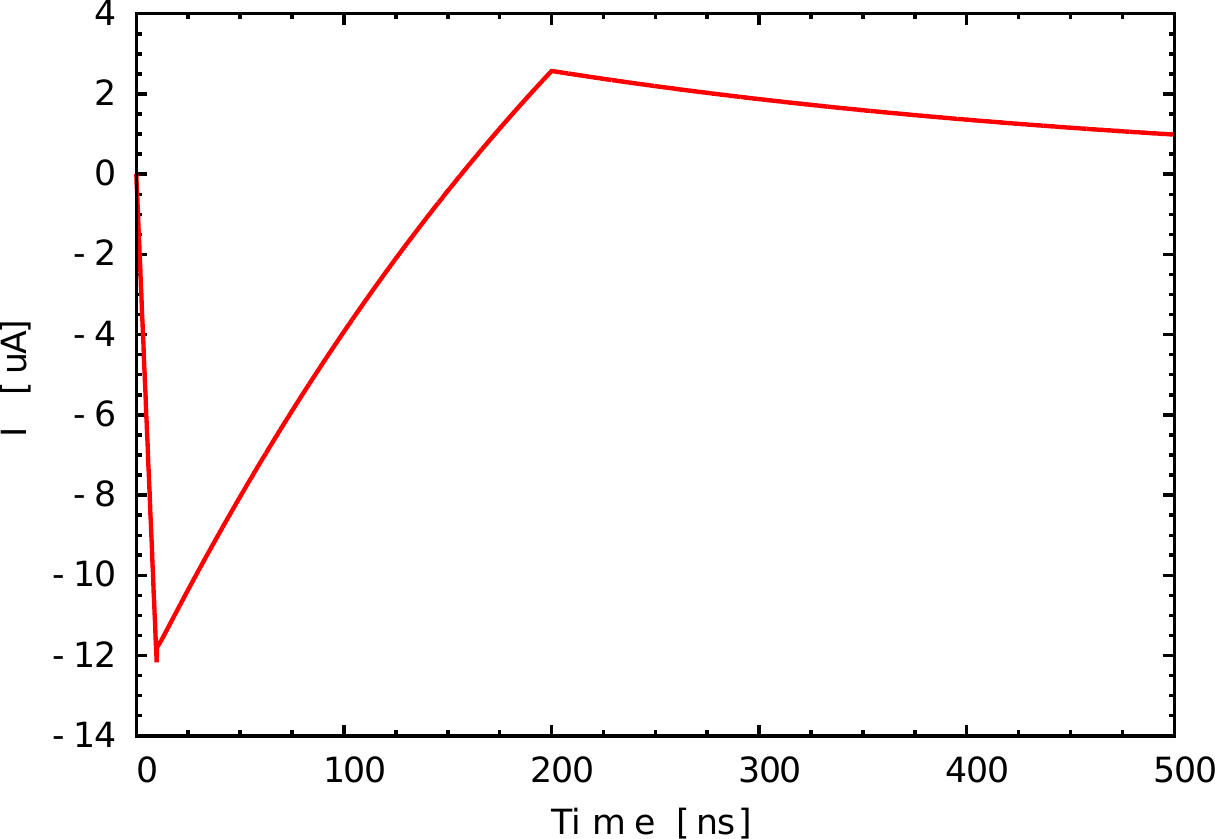}
 \includegraphics[scale=0.6]{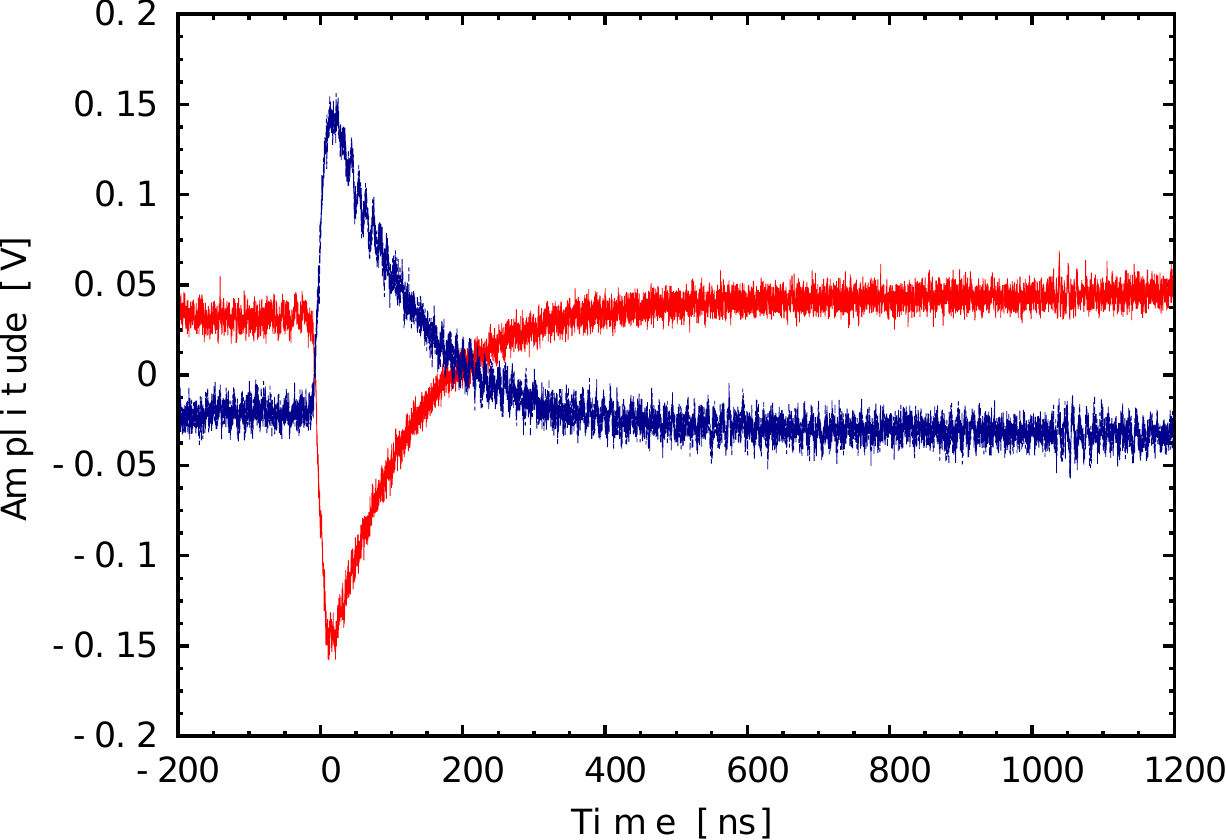}
 \caption{(color on-line) Upper panel: Preamplifier input current signal reproduced with the help of Qucs package. Lower panel: Measured output signal from the preamplifier. Blue and red lines denote positive and negative
output signals.}
\label{inppreamp}
\end{center}
\end{figure}

Signal readout from both ends of the relatively low resistance wire requires application of dedicated preamplifiers.
 Low resistance wires force the use of fast preamplifiers with low input impedance since the change of signal amplitudes
 and thus position resolution depends on the input impedance. The lower the input impedance, the higher the voltage difference
 that is registered at the wire ends. The preamplifiers work in current-mode with the input impedance below 5 ohms and the
 input stage bandwidth exceeding 300 MHz. The preamplifier circuit is shown in Fig.~\ref{preamp}. 
The first stage of the preamplifier is a current-voltage converter with two fast bipolar transistors. The voltage pulse is fed to the high-pass filter which cuts off
 the DC bias and the slow varying components of the signal. The first filter is followed by a low-pass filter integrating the pulse
 with 100 ns time constant. The signal is then amplified and transmitted to the differential amplifier used to drive the transmission line.
 Differential signal transmission increases the external noise immunity when unshielded twisted pair ribbon cables are used to connect
 the preamplifier outputs to the inputs of the analog module. An example of input and output signals from a preamplifier is shown in Fig.~\ref{inppreamp}. 
 The input current signal was reproduced with the help of Qucs package~\cite{9}.
Using bipolar transistors increases the preamplifier immunity to possible discharges in the gas chamber. Adding more robust input protection
 (e.g. a series resistor and protection diodes) was abandoned as it would increase the input noise and impedance thus worsening the charge division resolution.
 However, accidental discharges cannot be avoided completely and the need for replacement of damaged preamplifiers was taken into account in the design. 
The individual preamplifiers are arranged as small piggyback PCB cards mounted directly on the wire frames (Fig.~\ref{pcbpre}). They can be replaced easily without
 unplugging cables or any other major intervention in the setup. Additionally, such a solution minimizes the length of the unshielded connection between the
 wire end and the preamplifier input.
\subsection{Analog cards}
\begin{figure}[!h]
\centering
 \includegraphics[scale=0.45]{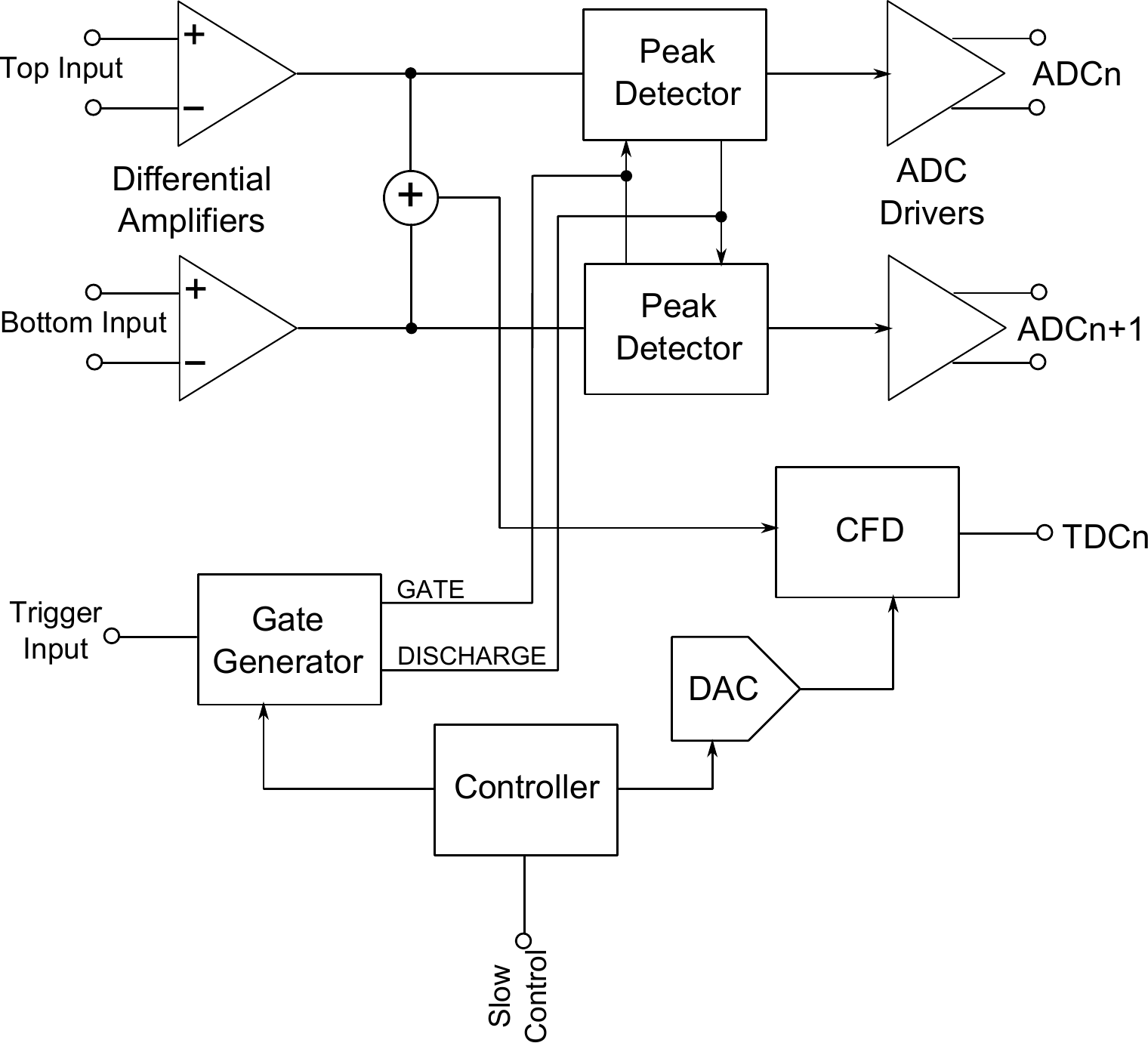}
 \caption{Block diagram of the analog board.}
\label{analog}
\end{figure}

The block diagram of the analog circuit is presented in Fig.~\ref{analog}.

\begin{figure}[t]
\centering
\includegraphics[scale=0.28]{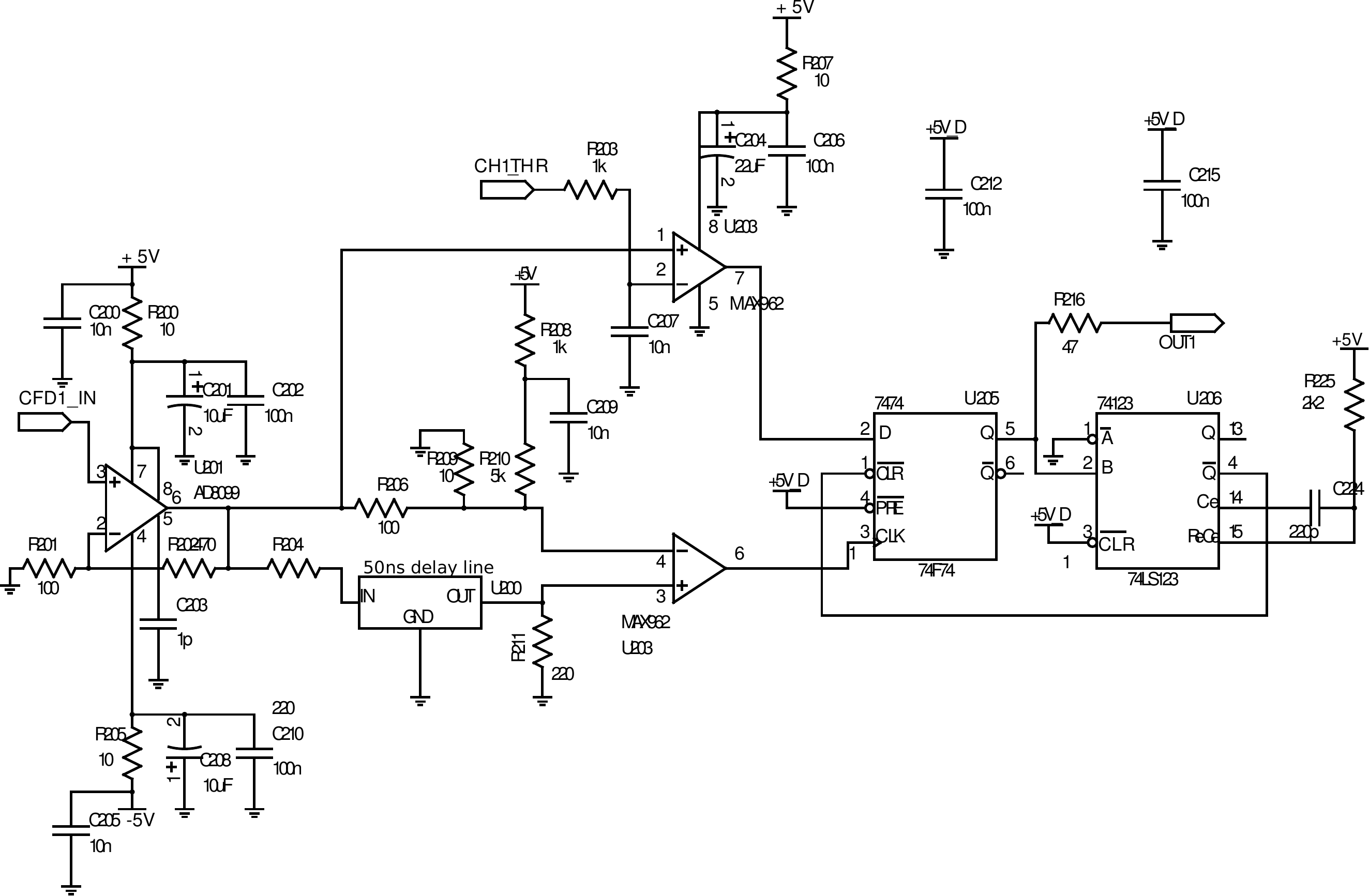}
 \caption{Schematic of the Constant Fraction Discriminator.}
\label{cfd}
\end{figure}

Differential signals from the preamplifiers are processed by the analog circuit. 
The signals from both wire ends are split into two branches. In the first branch, signals from both ends are added and fed to the CFD which produces
 a TTL pulse for the time-to-digital converter (TDC). A delay correction prior to the summing is not needed unless it is of the order of 1 ns 
(corresponding to about 20 cm cable length) – significantly less than the rise time of real signals from the MWDC. Care is taken to assure the equal
 length of the connecting cables. If the sum of the analog signals is higher than the set value of the CFD threshold, the STOP signal for the TDC is generated.
 The schematic of the CFD is presented in Fig.~\ref{cfd}.

\begin{figure}[!h]
\centering
 \includegraphics[scale=0.33]{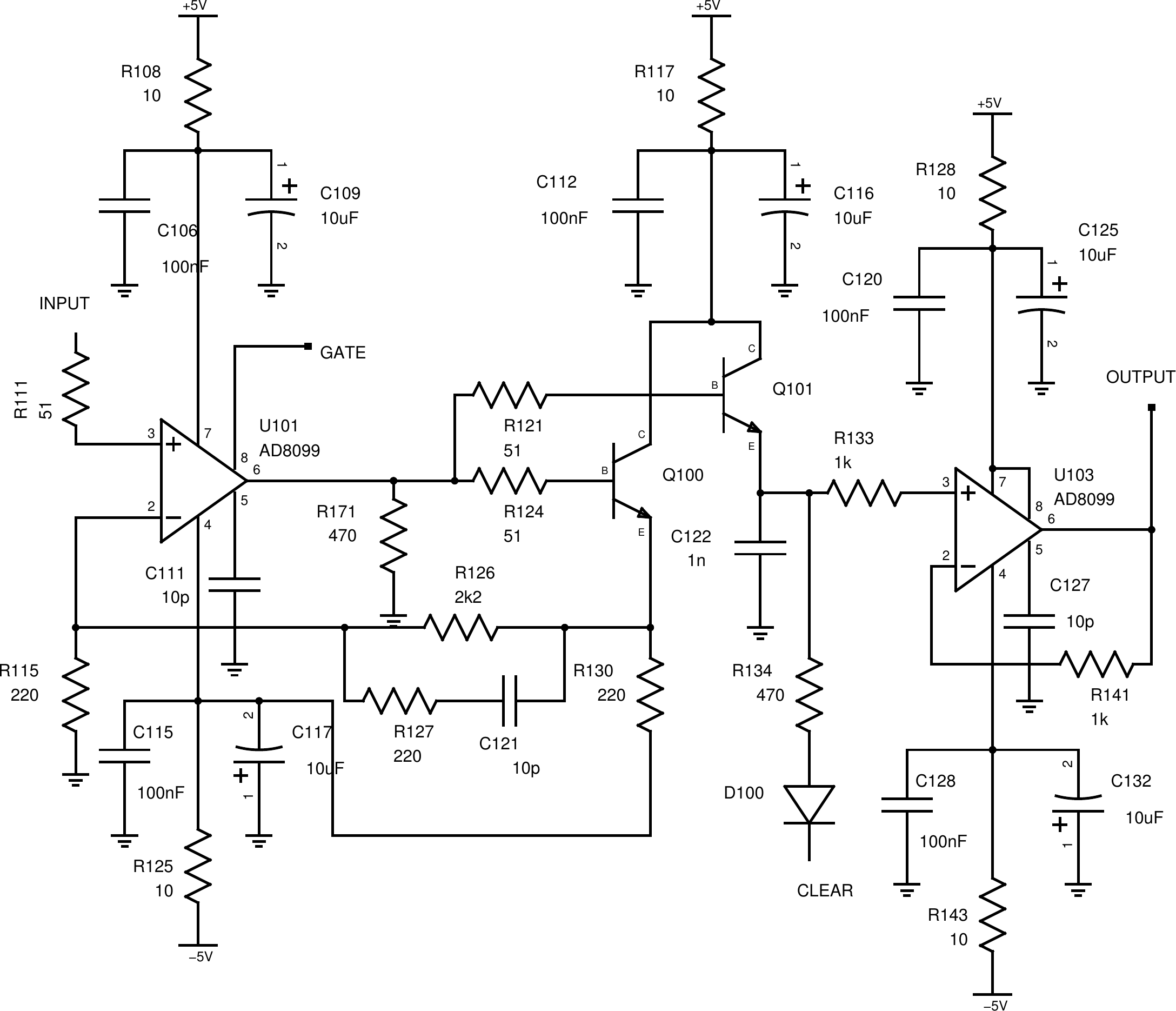}
 \caption{Schematic of the peak-hold detector.}
\label{phd}
\end{figure}

In the second branch, both signals are transmitted to the fast peak-hold detectors, which are responsible for detection of the pulse amplitudes in a given gate time.
 Peak-hold detectors are used to stretch the pulse to the length acceptable for the ADC converters.
 The schematic of the peak-hold detector is shown in Fig.~\ref{phd}. 

The gate and hold signals are produced by a programmable timing circuit
 located on the analog board which uses the TDC START signal (external trigger) as a time reference. The analog boards hosting the analog signal processing circuits
 are equipped with built-in controllers for setting the thresholds of the CFD as well as the gate and the hold timing. The set values are fed to the controller over
 a slow-control bus (RS485).

\subsection{Digital cards} 
\begin{figure}[t]
\centering
 \includegraphics[scale=0.39]{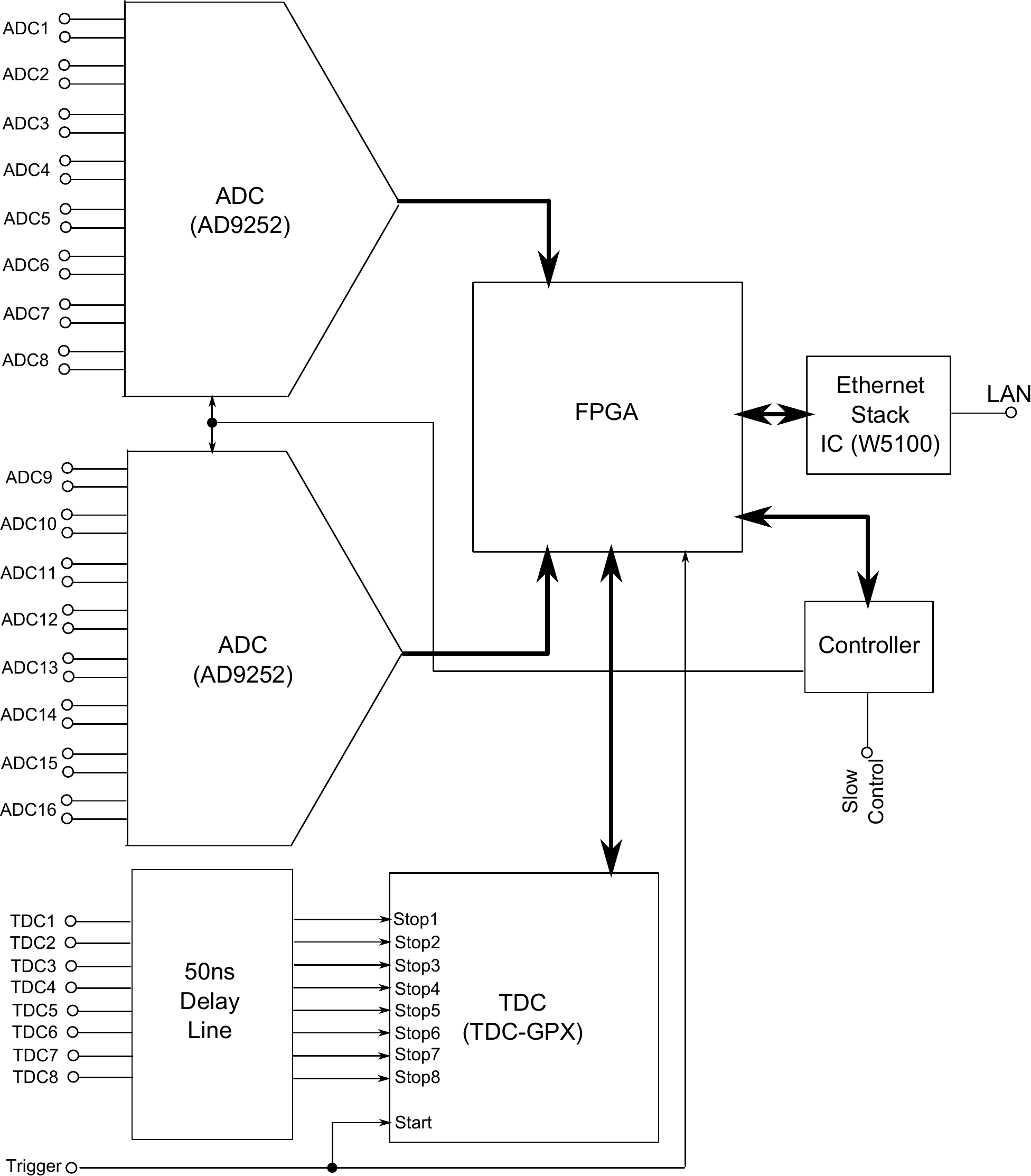}
 \caption{Block diagram of the digital board.}
\label{digitalb}
\end{figure}

The block diagram of the digital circuit is presented in Fig.~\ref{digitalb}. 
\begin{figure}[!h]
\centering
 \includegraphics[scale=0.5]{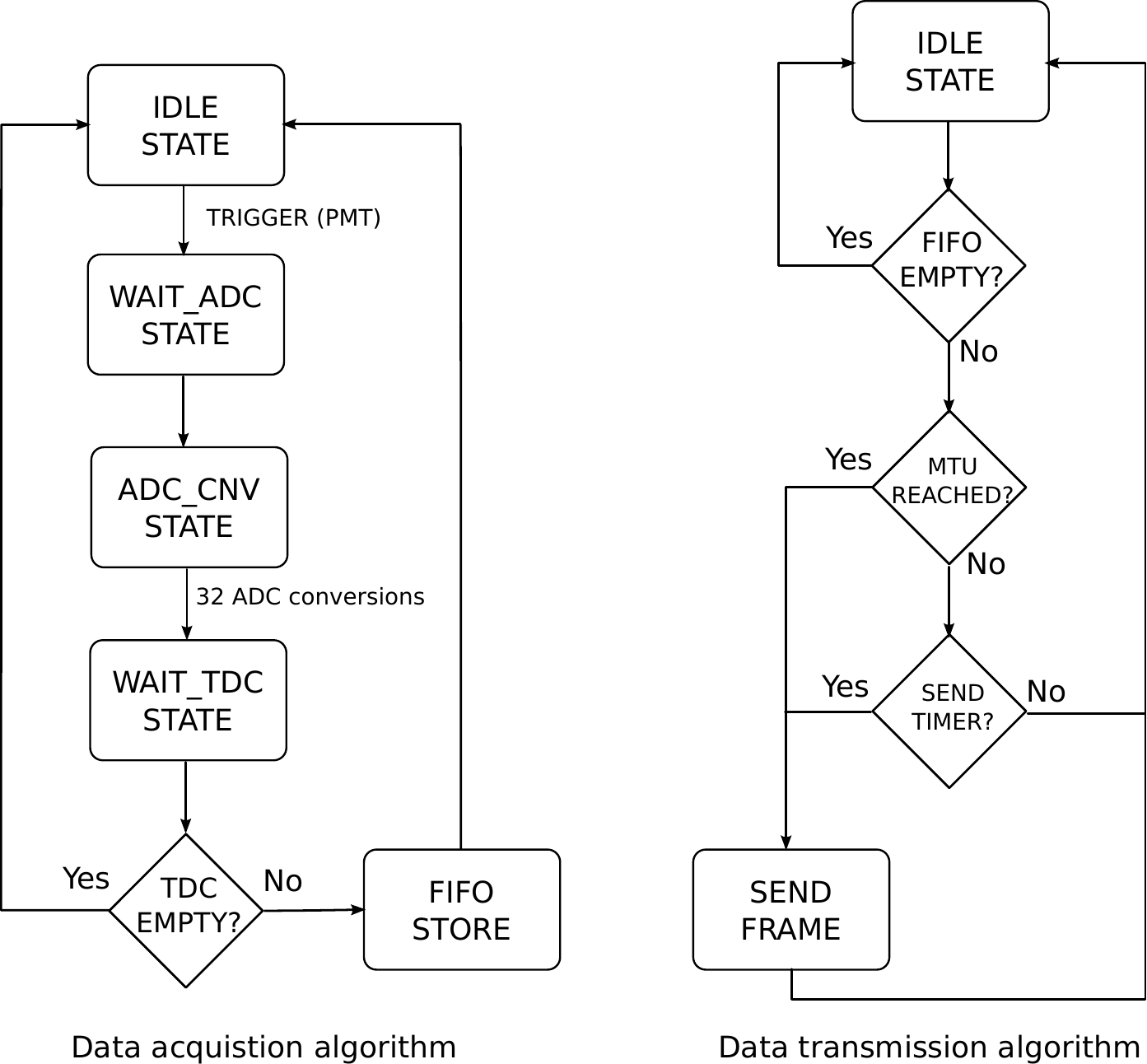}
 \caption{Flow diagram of the data acquisition algorithm (left panel) and of the data transmission algorithm (right panel).}
\label{daq_alg}
\end{figure}
The digital board consists of 16 channels of 14-bit ADCs (AD8099 chips made by Analog Devices) 
and a 8 channel TDC (TDC-GPX chip made by Acam Messelectronic Gmbh). Using 14-bit ADCs was dictated 
by the large dynamic range of the amplitudes of the signals from a gas detector operated in the proportional mode.
 The high bit ADC resolution increases the overall performance of the charge division method. The time resolution of
 the used TDC chip is significantly better than actually needed (83 ps binning resolution with the typical standard deviation
 of 77 ps, from the device datasheet). The board measures the delay between the common START (trigger) pulse and the individual
 STOP signals received from the constant fraction discriminators, as well as the amplitudes of the pulses delivered by the peak-hold detectors.
 The STOP signals are delayed by a 50 ns delay line in order to compensate for the delays introduced in the analog circuit generating the START signal.
 The ADCs sample the signals with 20 Msps sampling rate. The FPGA chip (Xilinx Spartan XC3S400) receives the data from the ADCs and the TDC, buffers them,
 provides zero-suppression and time-stamping, and formats the data frame as described in the next paragraph. The flow diagram of the acquisition algorithm 
and the data transmission algorithm are shown in Fig.~\ref{daq_alg}.
 The data frame is transmitted to the integrated Ethernet/UDP stack (Wiznet W5100) which is responsible for communication with the back-end PC hosting the acquisition
 and control software. 
 The digital board is also equipped with a built-in controller providing initialization and 
configuration of ADC, TDC and Ethernet chips. This controller is accessible via the slow-control bus. 

\subsection{Data formatter}
\begin{figure}[t]
\centering
 \includegraphics[scale=0.19]{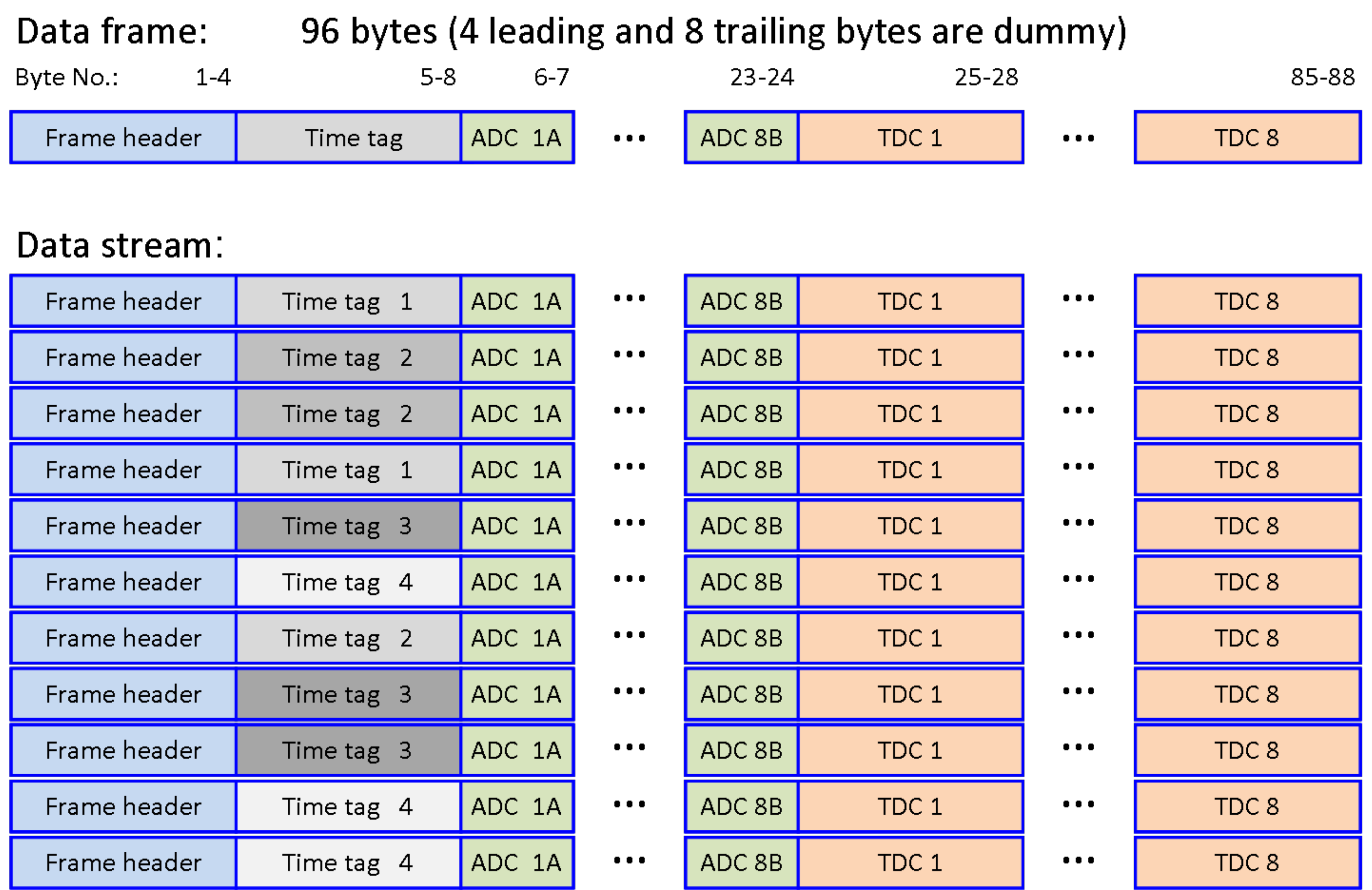}
 \caption{Data frame structure. The structures with the same time tag belong to one event.}
\label{formatter}
\end{figure}

The data are transferred to the computer in a form of UDP frames. 
The data formatter creates a frame for each recorded event. The frame length is a multiple of 96 bytes.
 The first part of the frame is a fixed header used for synchronization. Next follows the time tag (32 bit unsigned integer value)
 which allows the time synchronization between multiple modules. The third part of the frame consists of 16 ADC conversion values
 (16 bit unsigned integer values). The fourth part contains the TDC values - up to eight 24-bit unsigned integer values. Only the
 non-zero values of the TDC data are transmitted. The time tagging is used because the Ethernet protocol does not ensure the sequence
 of the data packet delivery. The time-tags are shared between modules and used by the data-logging software to restore the correct data order.
Fig.~\ref{formatter} shows the structure of the data frame. 

\subsection{Slow-control bus and interface}
\begin{figure}[!h]
\centering
 \includegraphics[scale=0.35]{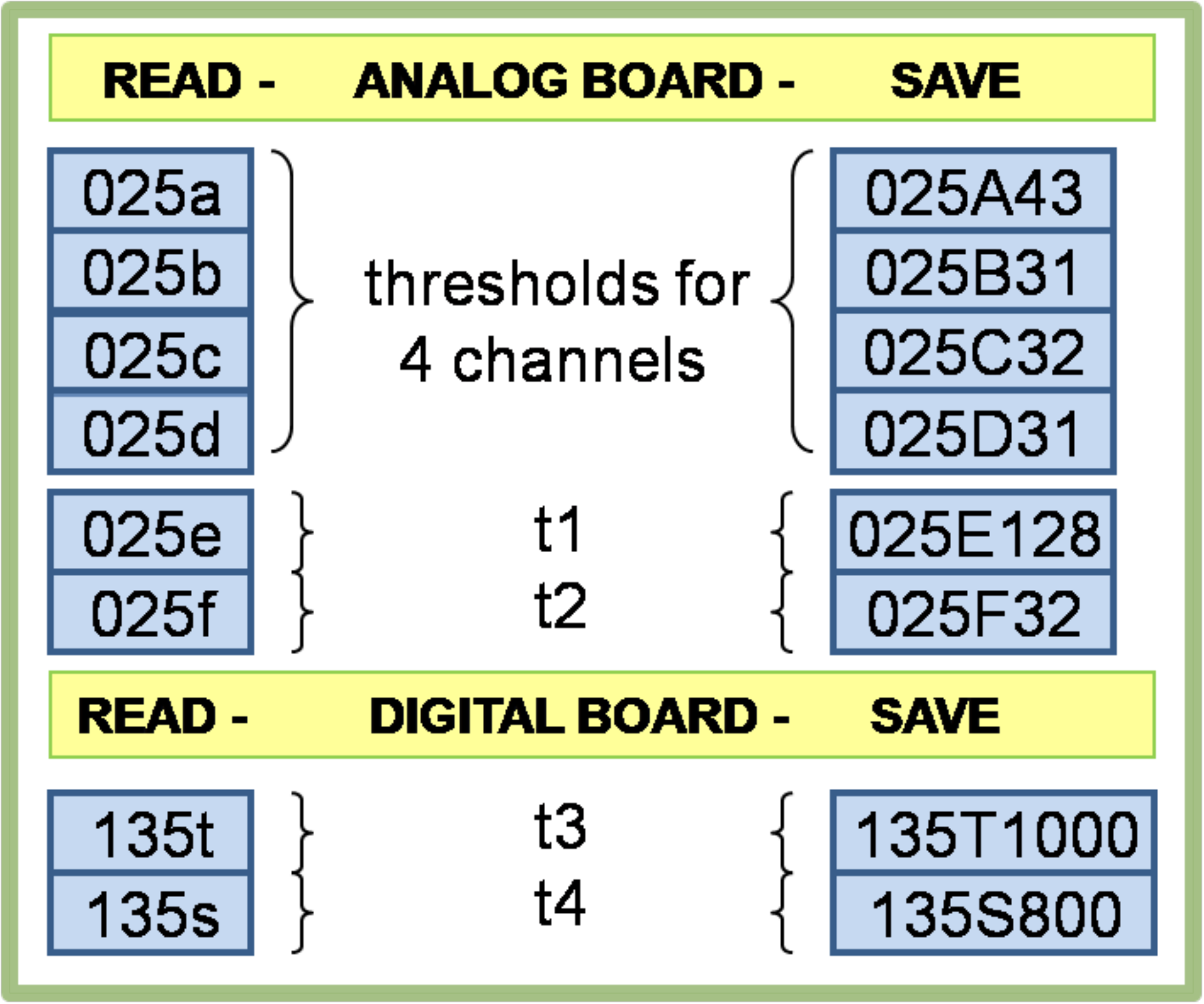}
 \caption{ASCII command format used for reading and saving values. The command consists of the controller address, the read commands for thresholds
(a,b,c,d), analog board timing ($t1$,$t2$) and digital board timing ($t3$,$t4$), where $t1$ is the gate width and 
$t2$ is the pulse hold time. For the digital board $t3$ denotes the maximum time for the incoming signal and $t4$ is the delay for the ADC. 
The second column presents the commands used for saving values to the controllers. The save commands use the same format but the uppercase option letters.}
\label{slow}
\end{figure}

\begin{figure}[!t]
\centering
 \includegraphics[scale=0.49]{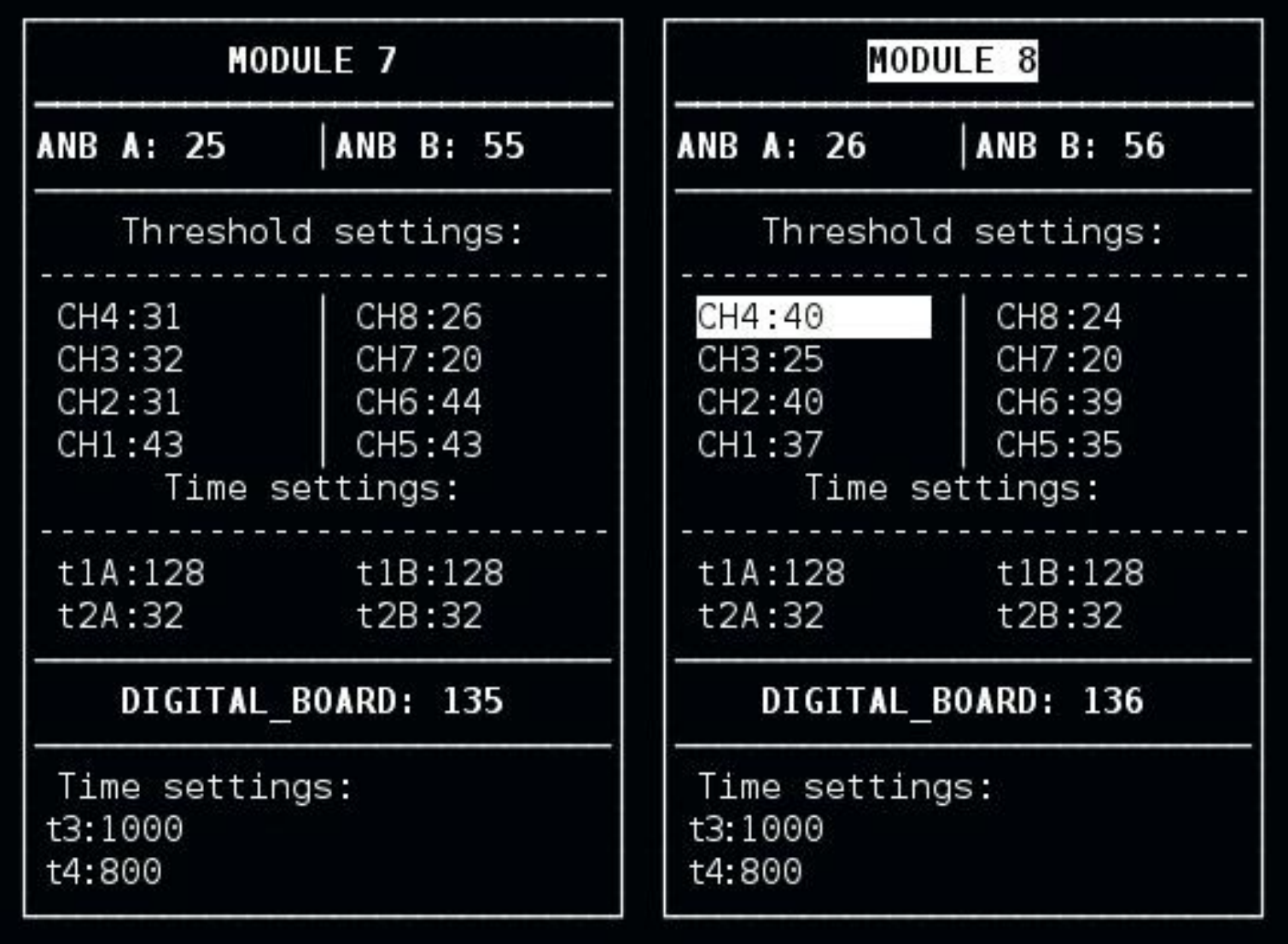}
 \caption{Example of the slow-control display. One panel corresponds to one module
consisting of two analog boards (ANB A and ANB B) and one digital board. Backlit represents the edited module data.
}
\label{scview}
\end{figure}

The slow-control interface allows settings the CFD thresholds and timing configuration parameters of the module.
 Each board in the module is equipped with its own controller connected to the slow-control bus. 
Individual addresses of the controllers are set with dip-switches installed on the boards. 
The controllers use a simple ASCII protocol via RS485 interface. In order to connect the bus
 to the acquisition computer a commercial RS485-USB converter is used. An example of the initialization
 data record necessary for establishing the communication over the slow-control bus is presented on Fig.~\ref{slow}.
 The slow-control parameter set is read back every second and refreshed on a computer display (Fig.~\ref{scview}).

\subsection{Data logging software}
\begin{figure}[!t]
\centering
 \includegraphics[scale=0.19]{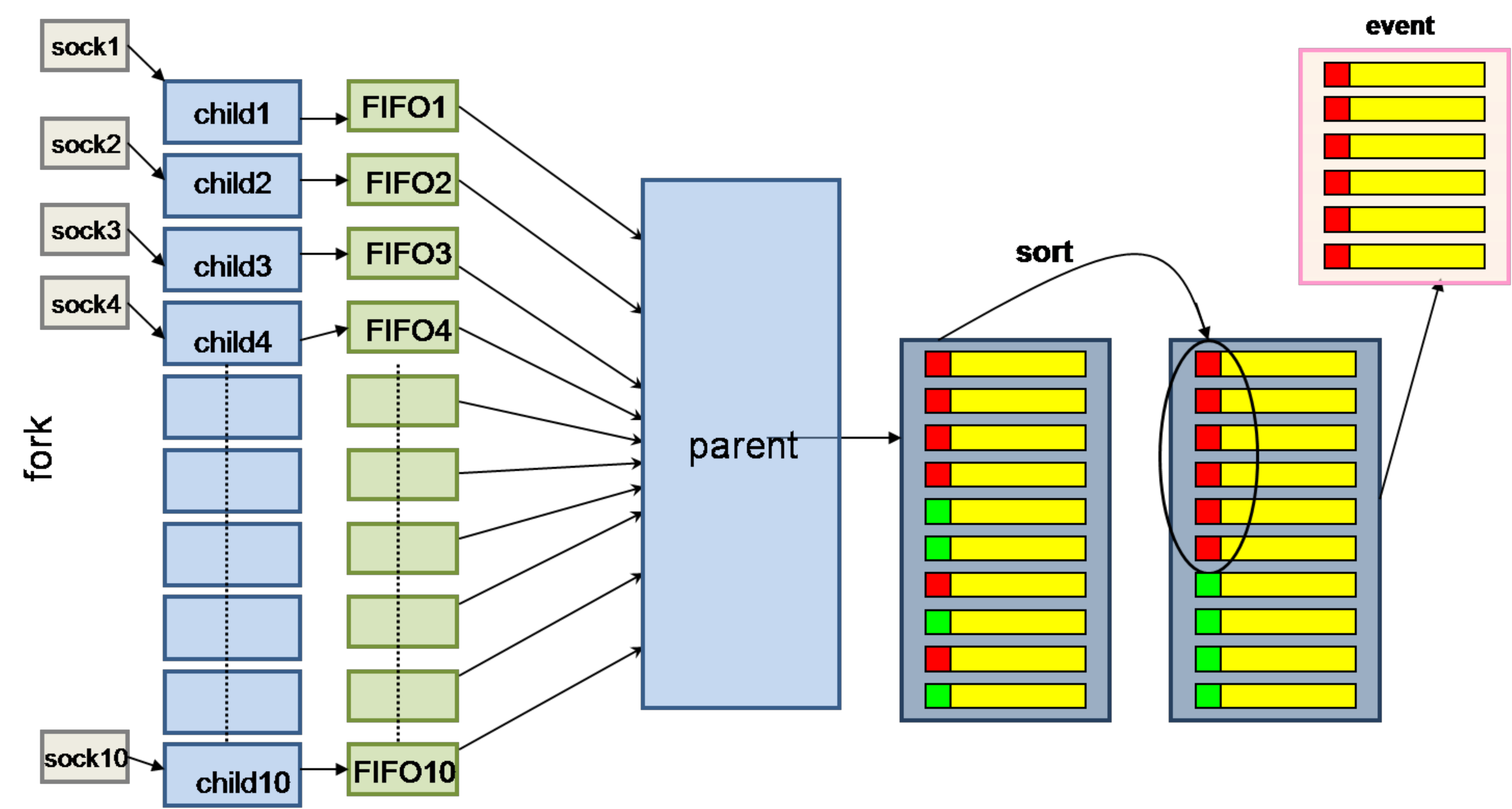}
 \caption{(color on-line) Data flow diagram for ten acquisition modules. The red (green) squares denote the same time tag.} 
\label{dls}
\end{figure}

The data logging software provides the parallel reception of data frames sent by the modules and the event building i.e. 
collection of data generated by the system upon a single hardware trigger signal and uniquely identified by a time tag.
 Each module is identified by its IP address and uses a specific UDP
port corresponding to this address. The architecture of the logging software is shown in Fig.~\ref{dls}.
The software is multi-threaded and consists of the main thread (the parent) that is responsible for sorting and recording events,
 and the child processes whose task is to receive data from the acquisition modules and to send them via data streams to the main thread.
 Upon initialization the program reads the configuration file with the addresses of the modules and opens sockets for all modules to communicate with them.
 The sockets listen to the different UDP ports correlated with the module address. The main loop creates a set of pipes (data streams) and receives
 the data structures from the child processes. The size of the frame is checked and followed by the pre-selection of cases. If the TDC value is greater than zero,
 the event is pushed to the FIFO queue of the child process. The event is then read from the FIFO by the parent process in which the data structures from all child
 processes are sorted (using the quick-sort algorithm) by the time tags and formatted as one complete event. This event will be identified with a particle crossing the
 chamber planes and directed for physics analysis. The received events are accessible for on-line histogramming analysis by means of a shared memory mechanism and 
simultaneously saved on a hard disk for later off-line analysis. 

 The cases with empty TDC conversion value (STOP signal below CFD threshold) are used to calculate the baseline offset for
 the ADC. In this method it is assumed that the lack of the TDC signal in a pair of channels (serving a particular sense wire) means that the cell did not fire so the
 corresponding ADCs deliver the offset values. This in-flight offset calculation method can be disabled and the fixed offsets can be introduced with the help of the
 slow-control interface. The baseline offsets subtracted (in flight) from the conversion value are appended to the data structures for each event so that the subtraction 
is reversible. Finally, the data structure consists of a time tag, module number, channel number, ADC1 value, ADC2 value, TDC value, ADC1 offset and ADC2 offset, respectively. 
It is worth mentioning that the amplitude thresholds which are needed for the histogram building were offset from the ADC baseline values such that only the electronic noise was eliminated. 
The ultimate adjustment of the thresholds was postponed to the detector efficiency tuning phase. It will depend on the gas mixture composition and pressure as well as on the operation voltage.
\section{Test results}
\begin{figure}[!t]
\centering
 \includegraphics[scale=0.54]{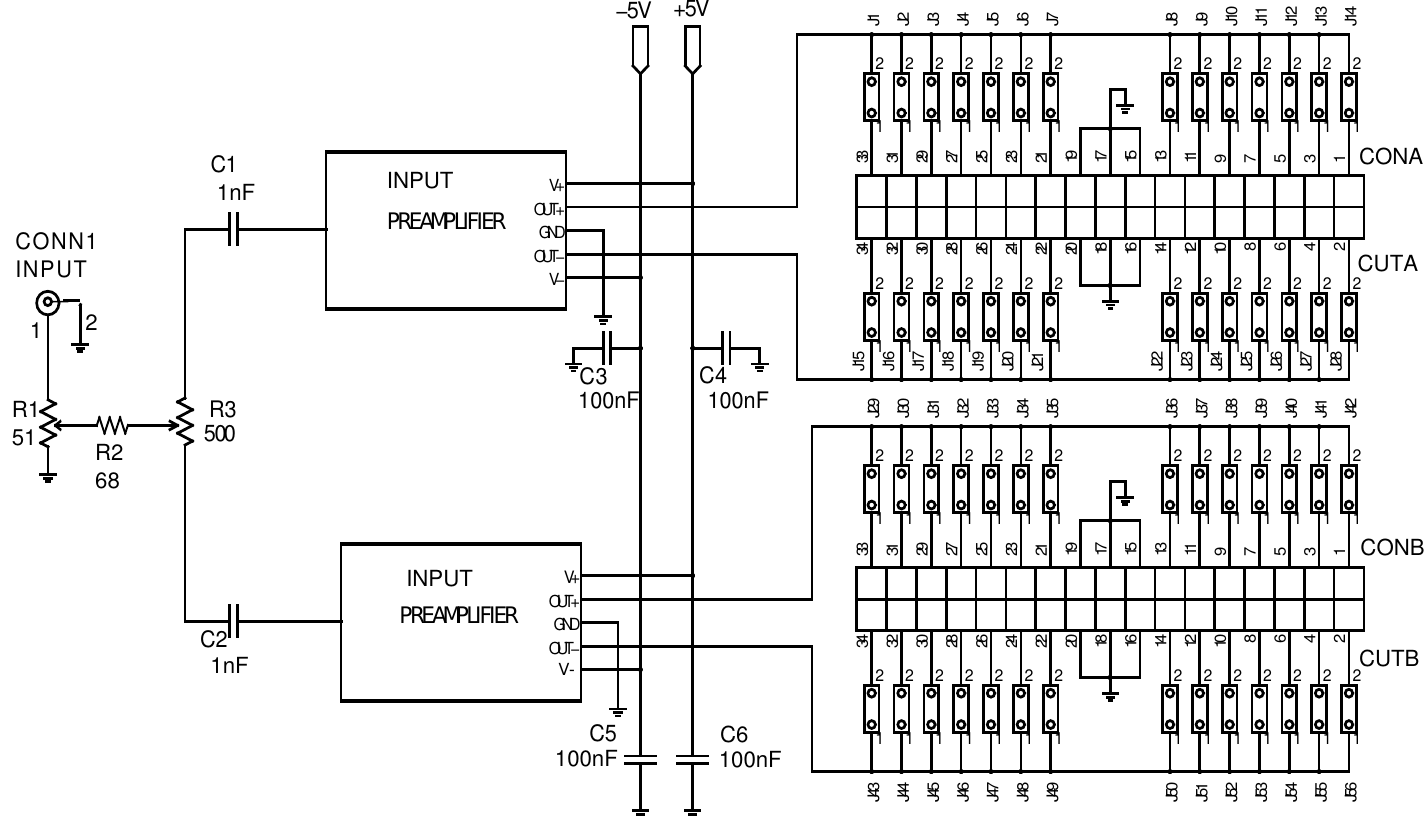}
\caption{Schematic of the dedicated tester for simulation of the signal wire.}
\label{tester}
\end{figure}

\begin{figure}[!t]
\centering
 \includegraphics[scale=0.24]{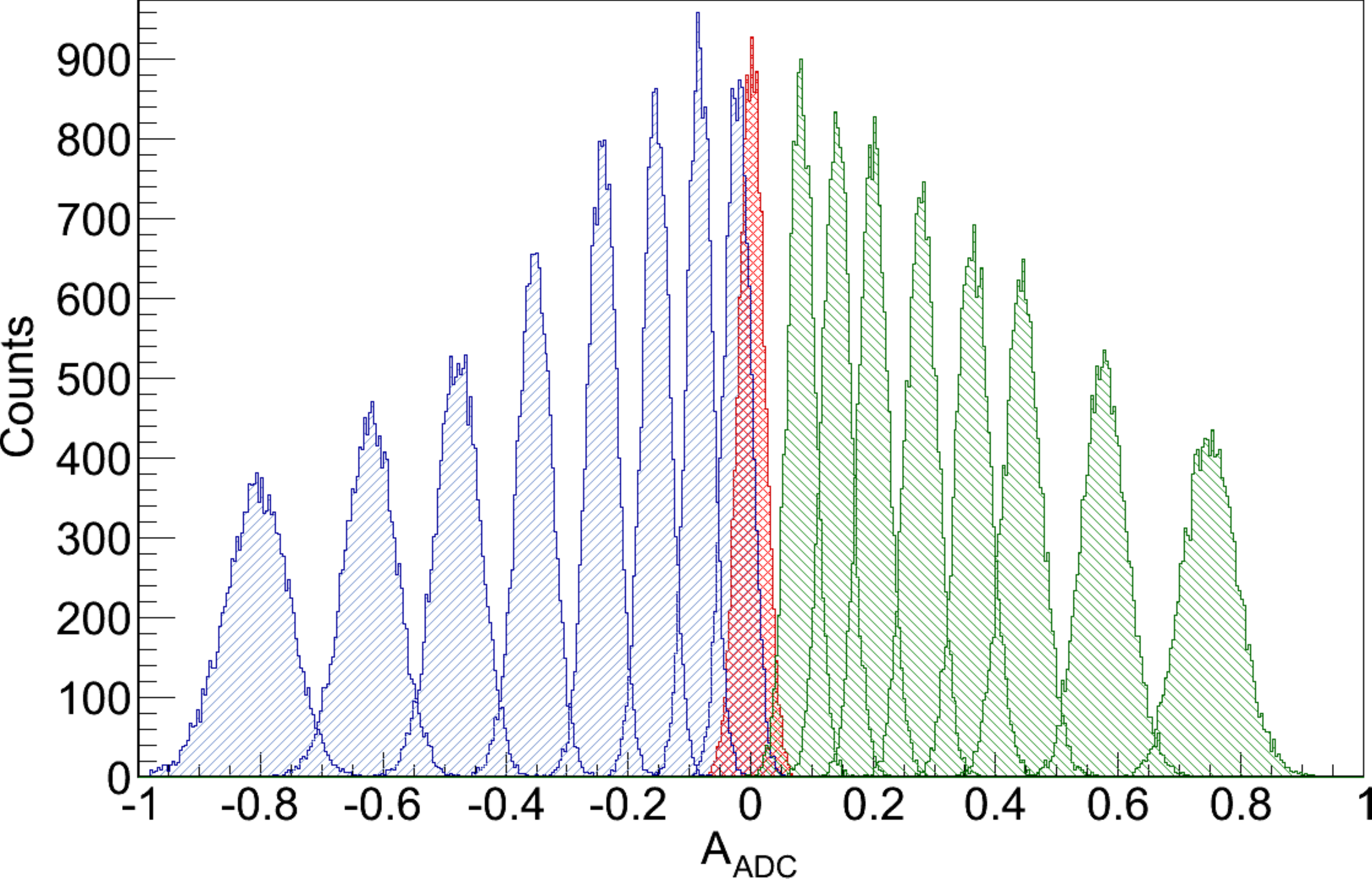}
 \caption{(color on-line) ADC asymmetry spectra taken for 12 different potentiometer settings. The red hashed peak corresponds to $A_{pot}$= 0.}
\label{adcall}
\end{figure}

\begin{figure}[!t]
\centering
\includegraphics[width=7.7cm,height=11cm]{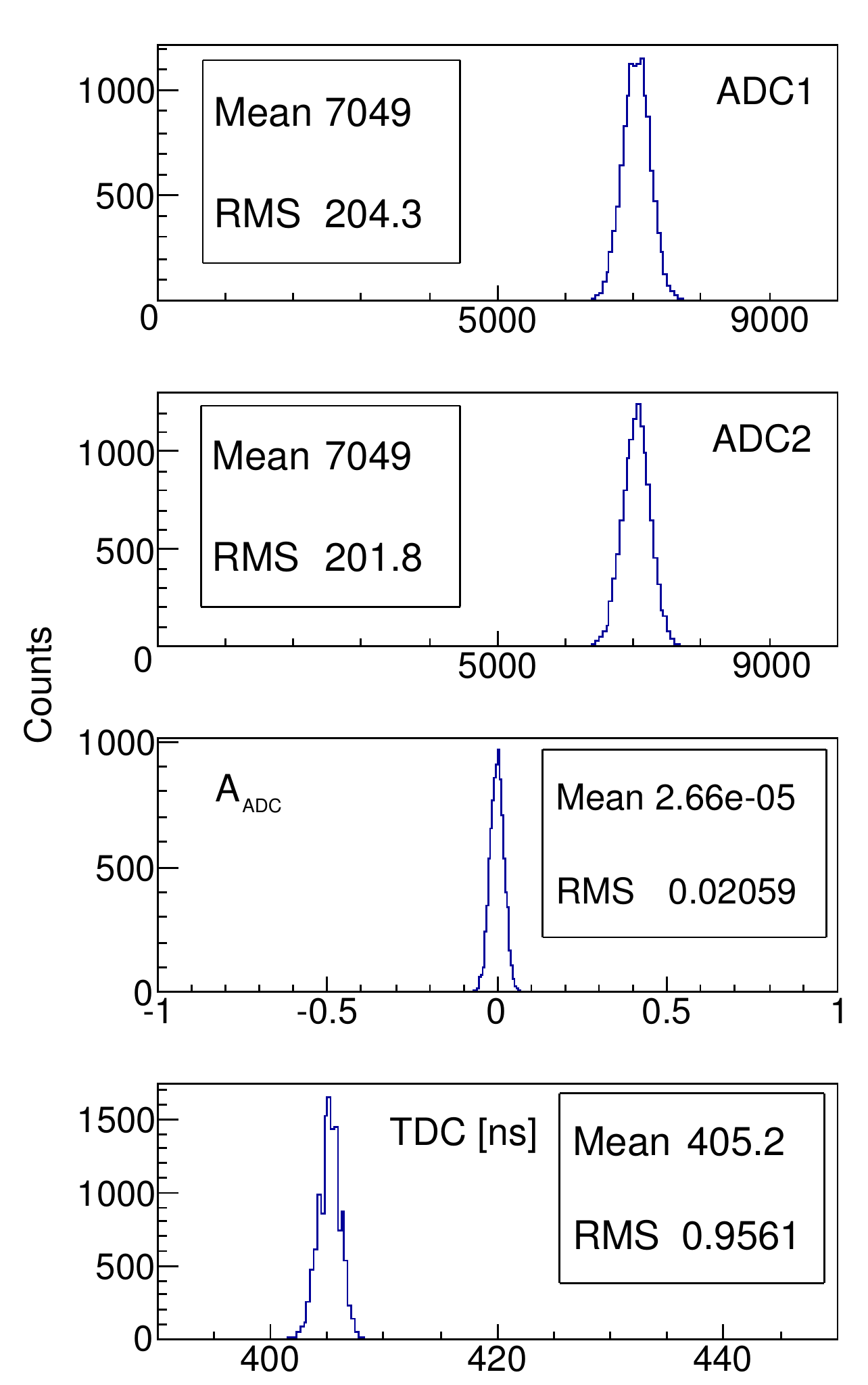}
\caption{Sample results for a single channel of one acquisition module. 
The two top plots present collected ADC values corresponding to both the upper part and the lower part of the wire. 
The two lower plots show the ADC asymmetry and converted TDC values in nanoseconds.}
\label{wzorzec}
\end{figure}

The described DAQ system was tested using a special tester simulating the wire chamber signals. 
The wire itself is replaced by a potentiometer with the range reflecting the real wire resistance (see Fig.~\ref{tester}). 
The tester utilizes two preamplifiers and signal cables connected to the selected module. A 400 mV high and 200 ns long simulator input is synchronized
 with a TTL signal triggering the system. The simulator input was fed by a triangle signal with 10 ns rise time and 200 ns fall time adjusted to the 
expected detector pulse shape obtained from the GARFIELD (Ref.~\cite{10}) simulation. 
The hit position corresponds to a unique potentiometer setting. The corresponding ADC asymmetry distributions are shown in Fig.~\ref{adcall}. 
The zero point defines the middle of the wire. The difference between the individual asymmetry spectra reflects the varying relation between the signal and
 noise amplitudes. 
In Fig.~\ref{wzorzec} sample results from one channel are presented. 

\begin{figure}[!t]
\centering
 \includegraphics[scale=0.62]{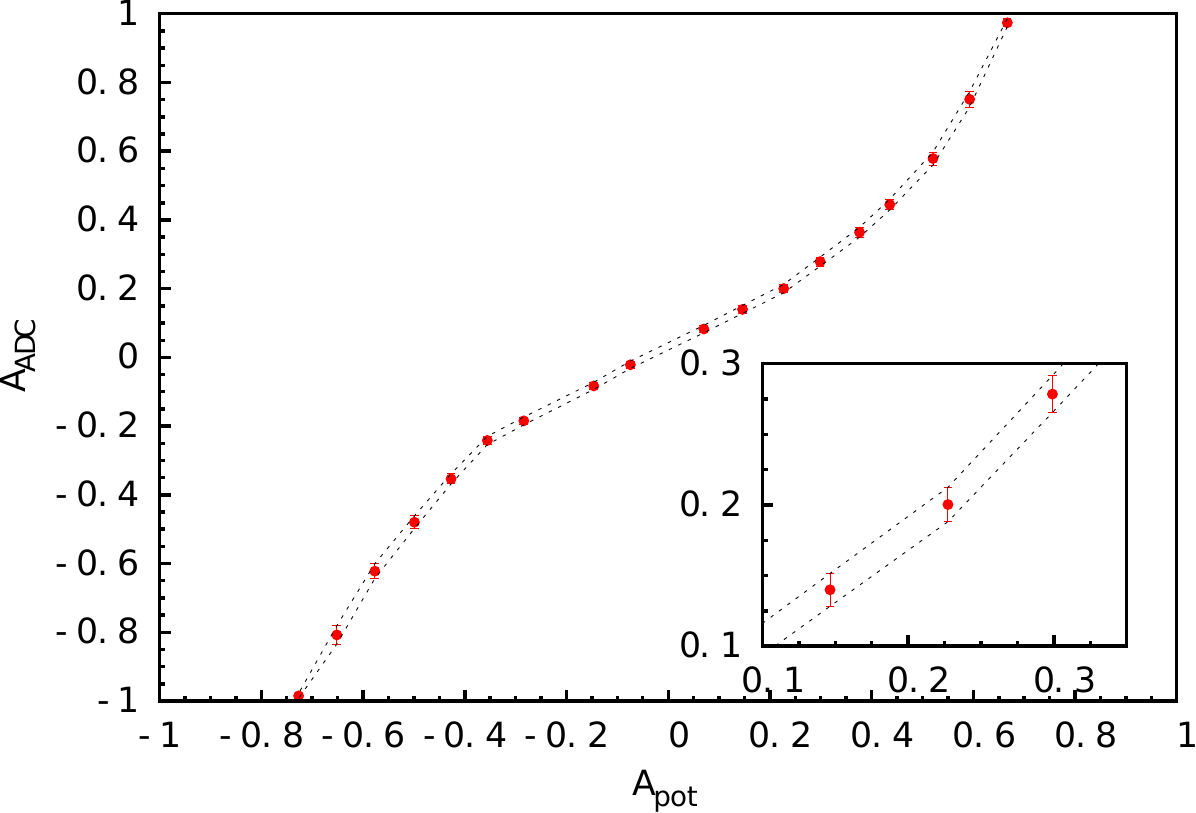}
 \caption{ADC pulse height asymmetry (centroids of the peaks from Fig.~\ref{adcall}) as a function of the resistance asymmetry $A_{pot}$ (Eqn. 2). 
The insert is a zoomed part of the graph showing the error bars equal to $\pm\, 1\sigma$ of the peak distributions plotted in Fig.~\ref{adcall}. 
Dotted lines interpolate the error bar ends.}
\label{funadca}
\end{figure}

\begin{figure}[t]
\centering
 \includegraphics[scale=0.59]{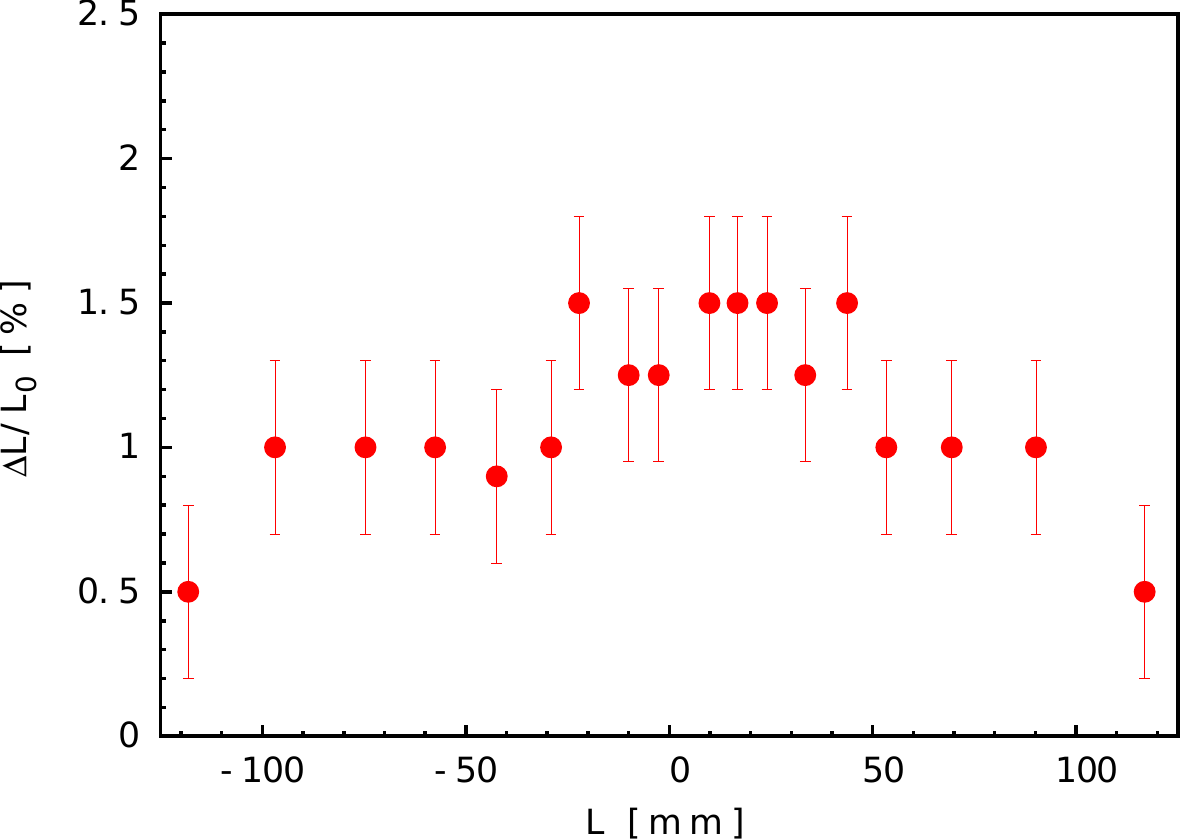}
 \caption{Position resolution $\Delta L/L_0$ expressed in units of length of the wire deduced from Fig.~\ref{funadca}. A 500 ohm wire resistance was assumed
 which corresponds to a 25 $\mu$m NiCr wire of $L_0$= 240 mm length. The uncertainty is connected with the interpolation procedure (see text).}
\label{funadcb}
\end{figure}

Fig.~\ref{funadca} shows a number of centroids the ADC pulse height asymmetry distributions 
\begin{equation}
 A_{ADC}=\frac{V_A - V_B}{V_A + V_B}
\end{equation}
acquired at different potentiometer asymmetry settings 
\begin{equation}
 A_{pot}=\frac{R_A-R_B}{R_A+R_B}
\end{equation}
where $V_A$ and $V_B$ are the ADC1, ADC2 amplitudes and $R_A$, $R_B$ are the resistances for selected potentiometer settings.  
 The results were obtained for varying resistance division corresponding to the charge collected at both wire ends. The obtained ADC asymmetry is a monotonic 
function of the resistance asymmetry. The centroids of these distributions are drawn with 1$\sigma$ error bars in Fig.~\ref{funadca}. The polynomial function
 fitted to the centroids represents the position calibration. Drawing the error band allows the extraction of the position resolution as shown in Fig.~\ref{funadcb}. 
The uncertainty of the relative position resolution is connected with the interpolation procedure.
It shows that the charge division method determines the position with a resolution between 1.2 and 3.6 mm for a 240 mm long wire. 
When feeding a given preamplifier pair with the simulator signals the noise distributions on the neighboring channels were acquired as well. 
Neither a baseline change nor an increase of noise were observed, meaning
 that the electronic crosstalk is negligible. This does not assure there is no crosstalk when the electronic system is attached to the gas detector. 
Inductive crosstalk between the wires depends on the configuration and on the
 operating conditions and is beyond the scope of this paper.

\begin{figure}[!h]
\centering
\includegraphics[scale=0.6]{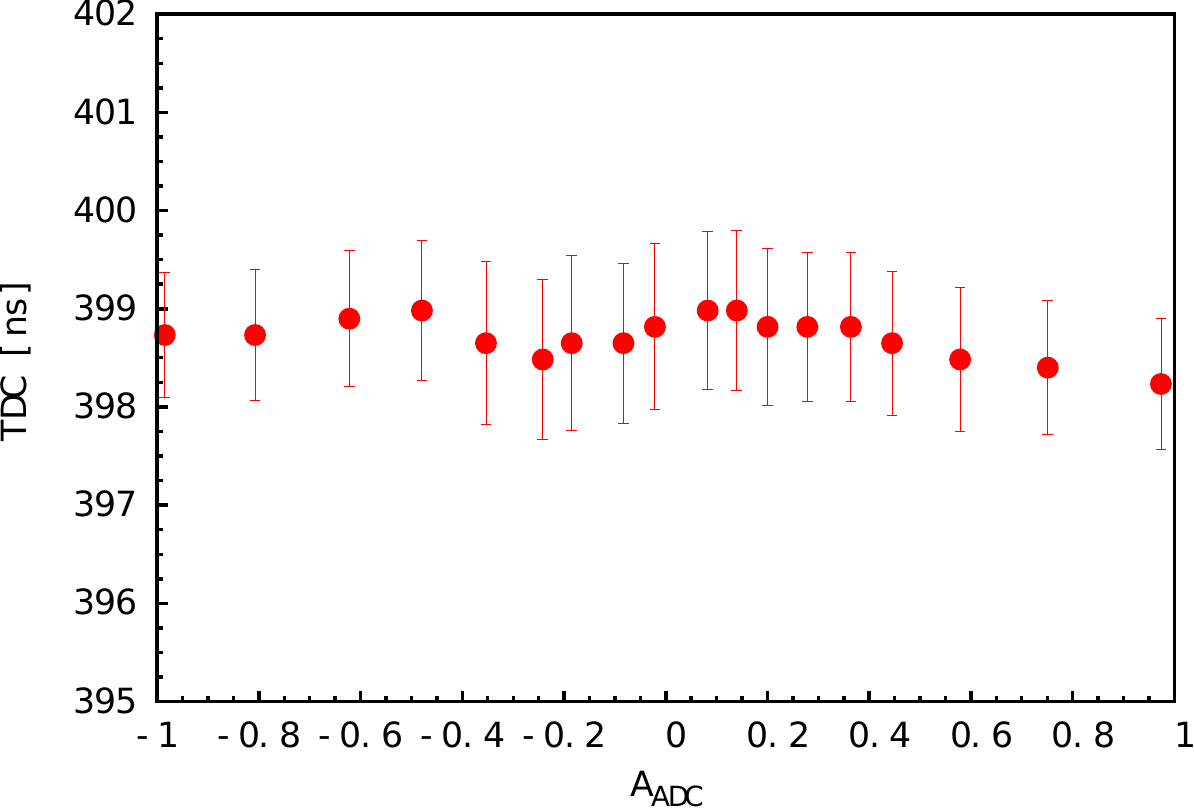}
 \caption{TDC values as a function of the ADC asymmetry. The error bars correspond to the standard deviation of the TDC histogram.}
\label{tdcconst}
\end{figure}

In the second test, the response of the TDC measurement was investigated as a function of the input signal asymmetry. Fig.~\ref{tdcconst} shows a typical example. 
The impact of the charge division on the TDC measurement is found to be less than 1 ns in the entire 
charge asymmetry range. For the application in view, a 1 ns drift time corresponds to about 250 $\mu$m distance (from GARFIELD simulations). 

The maximum throughput of the system reaches 15 kHz per channel (120 kHz per module) and is limited by the transmission time of 60 $\mu$s needed for a single event.
\section{Conclusions and outlook}
The front-end electronics and DAQ system described here was designed to be used with a special multi-wire drift chamber 
for tracking low energy electrons from nuclear $\beta$ decays. It incorporates both the drift time and charge division measurements allowing
 for efficient 3D determination of the electron tracks with very few and only parallel sense wires. The performed tests show that the electronic
 contribution to the drift time measurement uncertainty is less than 1 ns corresponding to about 250 $\mu$m position uncertainty at the expected electron
 drift velocities. Such result is satisfactory since in the planned experiment the track position resolution will be dominated by the electron angular
 straggling effects in the gas as shown in Monte Carlo simulations and confirmed in the small prototype test described in Ref.~\cite{8}. 
The necessary spatial resolution of the electron track position determined from the drift time need not to be better than 500 $\mu$m. The uncertainty of the track
 position obtained from the charge division measurement varies from 0.5\% at wire ends to 1.5\% in the center corresponding to 1.2 and 3.6 mm, 
respectively, for 24 cm long wires (25 $\mu$m NiCr). This result is sufficient for the application in view: identification of the 
cell sequence passed by an electron spiraling in an axial magnetic field and resolving possible double track ambiguities appearing in the 2D projection.

\end{document}